\documentclass[conference]{IEEEtran}
\IEEEoverridecommandlockouts

\usepackage{cite}
\usepackage{amsmath,amssymb,amsfonts}
\usepackage{algorithmic}
\usepackage{wrapfig}
\usepackage{algorithm}
\usepackage{graphicx}
\usepackage{textcomp}
\usepackage{xcolor}
\usepackage{cite}

\def\BibTeX{{\rm B\kern-.05em{\sc i\kern-.025em b}\kern-.08em
    T\kern-.1667em\lower.7ex\hbox{E}\kern-.125emX}}

\begin{document}

\title{CNN-based Speed Detection Algorithm for Walking and Running using Wrist-worn Wearable Sensors}

\author{

\IEEEauthorblockN{ Venkata Devesh Reddy Seethi}
\IEEEauthorblockA{\textit{Computer Science} \\
\textit{Northern Illinois University}\\
Dekalb, USA \\
devesh@niu.edu}
\and

\IEEEauthorblockN{ Pratool Bharti}
\IEEEauthorblockA{\textit{Computer Science} \\
\textit{Northern Illinois University}\\
Dekalb, USA \\
pbharti@niu.edu}

}

\maketitle

\begin{abstract}\label{section:Abstract}
In recent years, there have been a surge in ubiquitous technologies such as smartwatches and fitness trackers that can track the human physical activities effortlessly. These devices have enabled common citizens to track their physical fitness and encourage them to lead a healthy lifestyle. Among various exercises, walking and running are the most common ones people do in everyday life, either through commute, exercise, or doing household chores. If done at the right intensity, walking and running are sufficient enough to help individual reach the fitness and weight-loss goals. Therefore, it is important to measure walking/ running speed to estimate the burned calories along with preventing them from the risk of soreness, injury, and burnout. Existing wearable technologies use GPS sensor to measure the speed which is highly energy inefficient and does not work well indoors. In this paper, we design, implement and evaluate a convolutional neural network based algorithm that leverages accelerometer and gyroscope sensory data from the wrist-worn device to detect the speed with high precision. Data from $15$ participants were collected while they were walking/running at different speeds on a treadmill. Our speed detection algorithm achieved $4.2\%$ and $9.8\%$ MAPE (Mean Absolute Error Percentage) value using $70-15-15$ train-test-evaluation split  and leave-one-out cross-validation evaluation strategy respectively.
\end{abstract}


\section{Introduction}\label{section:Introduction}

The modern revolution of leading a healthy lifestyle is motivating people to incorporate physical exercises into their daily routine. Among various exercise options, walking and running are the most commonly practiced for fitness, health, and leisure \cite{Shipway2010}. These cardiac activities are also popular among the elderly population. Losing weight through exercise is directly associated with the intensity at which they are performed. For example, a person on average burns $1.5$ times more calories running at $5$ mph vs. $3$ mph \cite{Howley1974}. Similarly, walking at $2.5$ mph burns $2$ times more calories compare to $1.25$ mph \cite{sokhangoei2013investigating}. It makes fine-grained walking and running speed detection very important for people who want to keep track of their burned calories in any exercise session. Additionally, tracking the speed accurately can help people to understand their fitness levels and encourage them design custom exercise regimes and control their food intakes accordingly. In clinical settings, walking speed is monitored as an early symptom of dementia \cite{Welmer2014} and depression \cite{white2017association}.
    
Nevertheless, these facts and observations forge an absolute demand for a device that can measure the walking and running speed precisely in quickest time. For maximum outreach, the device should be accurate, pervasive, wearable and culturally acceptable among all the  age groups. Although, myriad of devices \cite{Bishop2010} are available commercially that aim for precise speed detection, their practical usage is limited. Majority of these devices \cite{Rampinini2015} are based on GPS that doesn't work well indoors. It can be problematic for people who live in colder places where walking outdoors is difficult. Vision based solutions are not pervasive \cite{Teerawat2019}, works as a stand-alone device and directly invades the user privacy. Other common problems with existing solutions are, either they are not accurate or highly energy inefficient. We will discuss them in detail in the related works section.  
    
    In this realm, a solution based on smartwatch/ fitness tracker can be proven promising considering its unobtrusive nature, pervasiveness, and convenient wearability. They are commercially available off the shelves at a modest price and have recently gained a lot of attention from the health and fitness communities. According to a recent study \cite{Ubrani2019}, International Data Corporation (IDC) reported an estimate of $69.3$ million smartwatches to ship in $2019$, and a total of $109.2$ million units to reach worldwide markets by $2023$ which shows that wearable based solution is future of fitness and healthcare. Further, a study in \cite{Santos-Gago2019} addresses the prevalence of wrist-worn wearables in the fitness industry, utilized by amateurs and professionals. Moreover, the study underpins the necessity to develop precise health and fitness applications by leveraging sensors embedded in the wearables.

    Modern smartwatches and fitness trackers are equipped with inertial measurement unit (IMU) sensors like accelerometer and gyroscope. These sensors are cheap, energy-efficient, and miniature enough to  fit in a smartwatch/ fitness tracker to capture the wrist movement precisely and swiftly. In comparison to GPS based solutions, IMU sensors work well at outdoors as well as indoors. In this paper, we have employed a wrist-worn tri-axial accelerometer and gyroscope sensors to precisely detect the walking and running speed swiftly. IMU signals can capture the variance between walking and running at different speeds (shown in Figure~\ref{fig:AccelerometerTri-Axial}). We designed and developed a deep convolutional neural network based regression model that takes input signals from wrist-worn accelerometer and gyroscope sensors, and predicts the walking and running speed. We extensively evaluated the algorithm and achieved favourable results presented in section \ref{section:Results}. In summary, our contributions in this paper are :
    
    \begin{itemize}
        \setlength\itemsep{-0.1ex}
        \item Designed and evaluated a convolutional neural network based regression model to estimate the walking and running speed from the wrist-worn accelerometer and gyroscope data 
        (Page $3-5$).
    
        \item Collected experimental data from $15$ subjects using Shimmer sensor \cite{burns2010shimmer} at a sampling rate of $51$ Hz. Data will be published on the author's Github account (Page $2$). 
        
        \item Presented a thorough evaluation of speed detector and step counting algorithms on leave-one-participant-out and train-evaluation-test dataset (Page $5$). 
    \end{itemize}

\section{Related Works}\label{section:relatedWork}

    Human activity recognition has been extensively explored by researchers by leveraging sensors based on smartphones, body-worn wearables, and wrist-worn wearables. In \cite{bharti2018human}, authors have classified $21$ complex in-home activities leveraging multi-modal sensors available in smartphones. Human activity classification in medical field can be used in circumspection of Parkinson's patients by detecting gait freezes using a waist \cite{Camps2018} or a wrist-worn IMU sensor \cite{Mazilu2016}. In clinical studies, the study in \cite{Bharti2017} used wrist-worn wearables to detect self-harming activities attempted by patients in psychiatric facilities. 
    
  In physical fitness and sports, sensing technologies can be used to gain insights on performance or the amount of work done. Each physical fitness activity is composed of a unique set of intricate motion artifacts. To capture these fitness-specific motion artifacts, several algorithms were proposed for counting repetitive exercises \cite{Morris2014}, assess volleyball skills \cite{wang2018volleyball}, and analyze stroke mechanics in swimming \cite{mooney2016inertial}. Similarly, algorithms proposed for walking-related activities were central to $3$ fundamental parameters of walking : step counts \cite{ngueleu2019design}, step length \cite{Hannink2018,martinelli2017probabilistic}, and walking speed \cite{Teerawat2019, Bishop2010, Zihajehzadeh2018, Mannini2011, Fasel2017, Soltani2019-mn, Bertschi2015-oe}. In previous research relating to speed detection, sensors were mounted either on shank \cite{Bishop2010}, thigh \cite{Mannini2011}, wrist \cite{Bertschi2015-oe, Zihajehzadeh2018}, etc. However, in performing walking related activities or daily house chores, sensor placed on wrist was found to be advantageous due to its high precision and recall rates \cite{atallah2011sensor}. The preciseness along with the prevalence of wrist-worn devices discussed in Section~\ref{section:Introduction} dissuades the use of body-worn sensors over wrist-worn devices. 

   In \cite{Bertschi2015-oe}, authors used a wrist-worn device with heuristic handcrafted features such as hand movement frequency, height, and weight of a person to predict the distance covered in a run. Study in \cite{Fasel2017,Soltani2019-mn} uses a personalized approach by constructing a relation between the stride length and cadence (i.e., number of steps in unit time) to find the speed of walking. Another approach to detect speed is using a combination of classification and regression models with two-stage support vector machines (SVMs) \cite{Mannini2011} or a combination of SVM and Gaussian random process (GPR) \cite{Zihajehzadeh2018}. However, heuristic and machine learning models are being replaced with the recent novel avenue of neural network research. The neural network architectures are used in applications relating to human activity recognition \cite{Ordonez2016-uz}, and predicting relative physical activity intensity of walking-related activities \cite{Chowdhury2019-to}. In this perspective, we demonstrate a precise running speed detection algorithm by leveraging deep convolutional neural networks and a wrist-worn sensor. Authors in \cite{wang2019speed} used a deep CNN with only a single accelerometer, and did not mention the number of participants involved and the type of wearable used in the data collection procedure. In addition, their model  achieved a mean error of $7\%-18\%$ for running and walking speeds while we had $4.2\%$ for train-test-evaluation split and $9.8\%$ for leave-one-participant-out (they didn't evaluate on this) cross validation. Our research displayed higher precision and a through evaluation by leaving one participant out at a time that depicts a practical use-case of a device where a person expects higher precision to plan their fitness regime.


\section{Data Collection and Preprocessing}
 \label{section:dataCollectionPreprocessing}
   
\begin{figure}[!ht]
    \includegraphics[width = \linewidth, keepaspectratio]{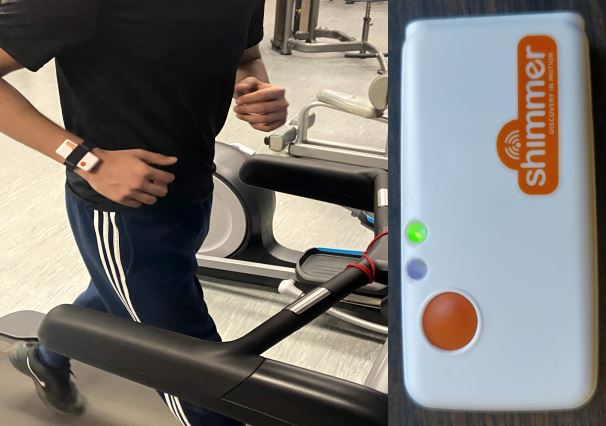}
    \caption{Participant wearing Shimmer device on the wrist while running at treadmill (left), Shimmer device (right) }
\label{fig:participantRunning}
\end{figure}

    In this study, we have employed Shimmer \cite{burns2010shimmer} IMU sensor as a wearable device for collecting the accelerometer and gyroscope data from the participants. Due to its small form factor, it can be easily worn on the wrist like a smartwatch (as shown in Figure \ref{fig:participantRunning}). Shimmer is widely used in the research community to study human body movements \cite{bevilacqua2018human, billeci2019autonomic}. This device is based on TinyOS firmware, with the core element as a low power MSP$430$ CPU with $24$ MHz clock rate that controls the device operation. The CPU has an integrated $16$-channel $12$-bit analog-to-digital converter which is used to constantly sample and capture tri-axial acceleration and rotational signals from the inbuilt accelerometer and gyroscope sensors respectively \cite{burns2010shimmer}. The accelerometer has two settings, low noise and wide range with ranges of $\pm 2$g and $\pm$ $16$g (where g is the gravitational acceleration) and the gyroscope has a range of $\pm$ $2000$ dps (degrees per second). In this study, low noise accelerations from accelerometer in the range of $\pm 2$g and angular velocities from gyroscope in the range $\pm$ $1000$ dps were collected where both of these sensors were sampled at $51$ Hz. During the experiment, sensory data from the device were streamed to an Android phone (OnePlus $5$T) via an inbuilt radio module  using the Shimmer Connect android application. Data from phone was later transferred to a server for post-processing and modeling.  

\subsection{Data Collection Procedure}
\label{section:dataCollection}

    We collected the sensory data from $15$ healthy adults ($8$ males and $7$ females) in this study. These participants belonged to the age group of $22\pm2$ years and had an average height of $168.5\pm13.5$ centimeters. The experimental protocol was approved by the Institutional Review Board (IRB) office at the Northern Illinois University. All participants signed the IRB consent forms and indicated that they were not fatigued and did not have any physical injuries or disabilities at the time of experiment.
    
    Before the experimental procedure,
    Shimmer was mounted on the participant's preferred wrist. Although, we intended to collect the data in the range of $3$ to $7$ mph, participants were allowed to set their own limit within our range. Each participant performed walking/ running on the treadmill for $45$ seconds session at a fixed speed. Speed was gradually increased by $0.5-0.7$ mph in every new session, based on the participant's preference. The experiment was carried out for $10-20$ minutes for each participant until they completed $9-16$ sessions at different speeds. In total, $205$ sessions were recorded from $15$ participants which comprised $9225$ seconds and $313650$ samples of accelerometer and gyroscope data from walking/ running. 

\vspace{-.75em}
\subsection{Ground Truth Collection}
\label{section:groundTruth}
\begin{figure}[!ht]
    \centering
    \includegraphics[width=\linewidth,height=4cm]{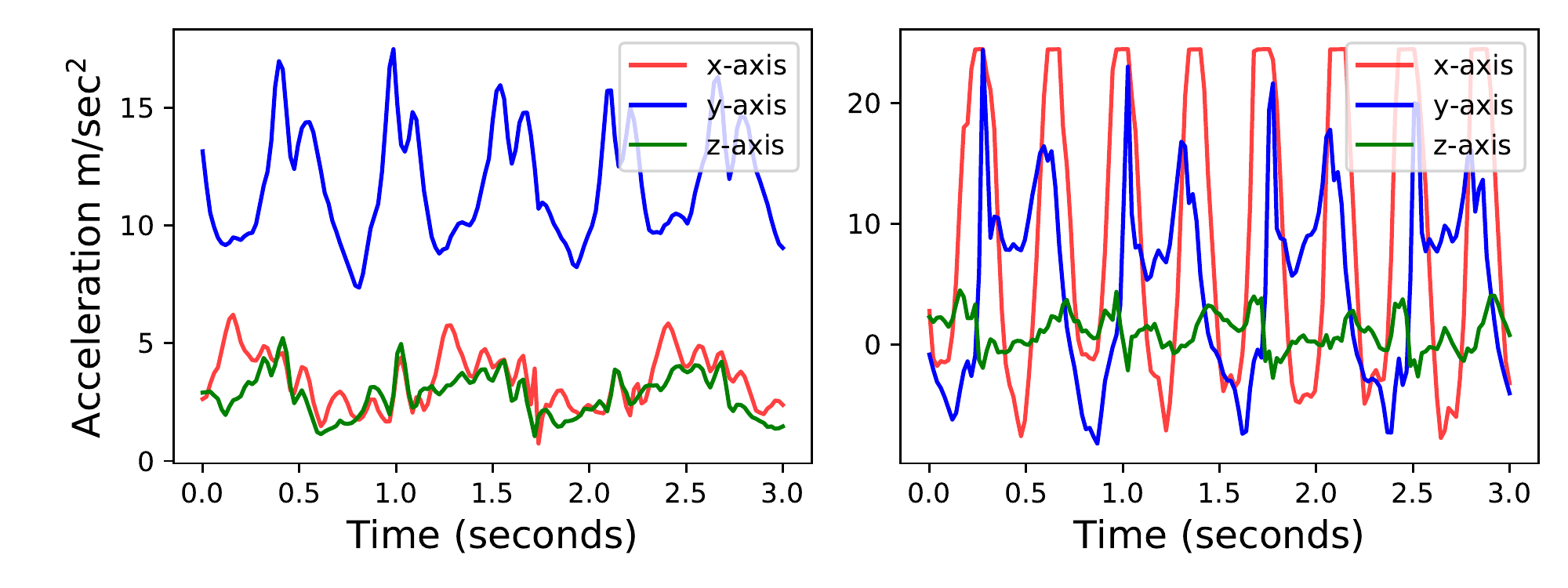}
    \caption{Tri-axial accelerations of a participant while 
                   (a) walking at $3$ mph,
                   (b) running at $7$ mph}
    \label{fig:AccelerometerTri-Axial}
    \end{figure}

    The experiment was moderated by the authors for streaming sensor data from Shimmer to the phone. Both sensors were sampled at $51$ Hz sampling rate. A new CSV (comma-separated values) file is created for each $45$ seconds session and tagged with a unique participant ID and their running speed. Sample raw accelerometer data from a participant at different speeds are shown in Figure~\ref{fig:AccelerometerTri-Axial}.


\vspace{-.5em}

\vspace{-0.5em}
\subsection{Data cleansing}
\label{section:dataCleansing}
    Data cleansing is an important aspect of data processing to remove the noise before further analysis. During the experiment, some participants were unable to run for $45$ seconds and, in such event, the readings were discarded to maintain the consistency in the size of data samples at each speed. Additionally, to discard the noise associated with the start and stop of each experiment session, the recorded data of $45$ seconds was trimmed by discarding $2.5$ seconds of data from the beginning and the end of recording, making each session effectively a length of $40$ seconds. 
    
    Like other sensors, IMU suffer from bias, alignment, and sensitivity errors that affect the accuracy of the signals produced by them. Shimmer provides calibration of sensory data to minimize these errors. We leveraged the calibrated value of tri-axial accelerometer and gyroscope data to train a convolutional neural network based algorithm to detect the speed.

\begin{figure}[!ht]
    \centering
    \includegraphics[width =\linewidth, height = 4cm]{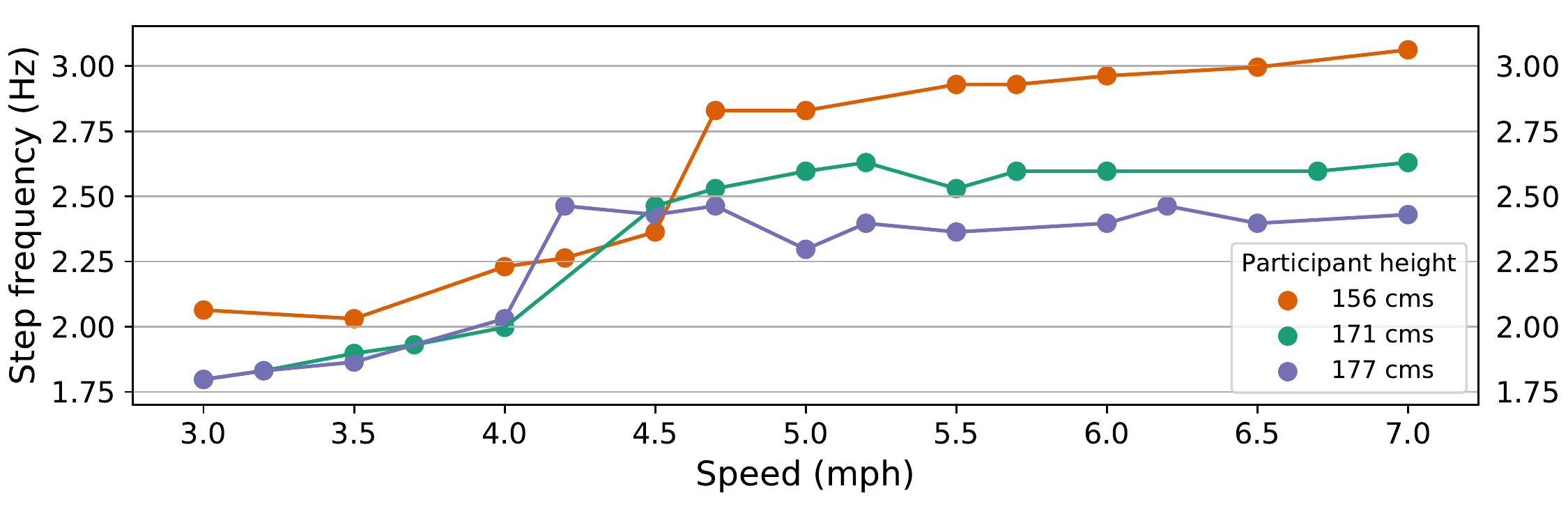}
    \caption{Step frequencies for three participants while walking/running at different speeds}
    \label{fig:StepFrequency}
\end{figure}
\vspace{-1em}
\begin{figure*}[!ht]
    \centering
    \includegraphics[scale=.9, keepaspectratio]{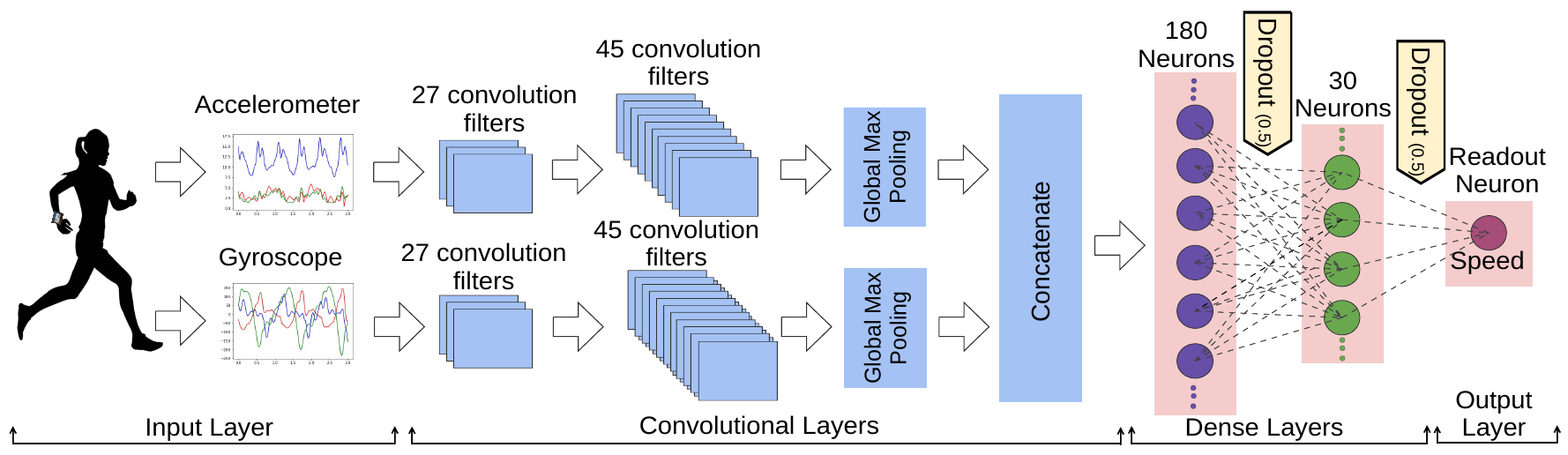}
    \caption{Architecture of CNN-based speed detection algorithm}
    \label{fig:speedArchitecture}
\end{figure*}

\section{Speed Estimation using Convolution Neural Networks}
\label{section:speedEstCNN}
Running or walking is a repetitive bipedal motion, where each strike of feet to the ground propels the person forward with a certain displacement. These displacements (step-length, $l$) along with the rate of strike of feet (step-frequency, $f$) are two critical components to determine the speed of walking and running. Speed $S$ can be calculated using formula $S$ = $l \times f$. $S$ can only increase if one or both of $l$ and $f$ increases. We derived the step-frequency $f$ by applying Fast Fourier transform (FFT) to the IMU signals\footnote{Due to space constraints, details of FTT based step-frequnecy computation is not provided.}. Fine-grained speed estimation using IMU signals is a difficult problem because although they can easily capture the variation in step-frequency, they fail to detect the change in step-lengths. However, in this study we observed that  step-frequency remains almost the same when speed increases within the walking or running activities as shown in the Figure \ref{fig:StepFrequency}. That means only step-length increase which is hard to capture using IMU sensors. In Figure \ref{fig:StepFrequency}, when a person increases their speed from $3.0$ mph to $4.0$ while walking, there is a negligible change in step-frequency observed. Similarly, when running speed changes from $5.0$ mph to $7.0$, the step-frequency curve is almost flat. Step-frequency only changes significantly when the transition in activities detected, i.e. - walking to running or running to walking. To overcome this issue, we trained several feature-based machine algorithms to estimate the speed but they didn't perform well. Finally, we designed a Convolutional Neural Network (CNN) based regressor architecture to achieve the desired results. CNN is a very popular neural network algorithm. It has achieved state-of-the-art results in various problem domains especially in computer vision. 




 \begin{algorithm}
 \caption{Algorithm for sliding windows}
 \label{label:slidingWindows}
\textbf{Input:} Dataset $data$, Frame size $F_{size}$, Overlap $ov\%$. \\
\textbf{Output:} Segmented data $Segmented\_sliding\_windows$.\\
\vspace{-1em}
\begin{algorithmic}[1]
\STATE $total_{sample-size}\gets418200$\\
\STATE  $Frame\gets 0$, $I_{s}\gets 0$, $I_{e}\gets F_{size}$ \\
\STATE $n\gets \dfrac{total_{sample-size} - F_{size}}{F_{size}\times (1 -ov\%)} + 1$ \\

\WHILE{$Frame \neq n$} 
      \STATE{$data_{slice} =  data[I_{s} : I_{e}]$} 
      \STATE{$Segmented\_sliding\_windows.append(data_{slice})$}
      \STATE{ $I_{s} = I_{e}$}
      \STATE{$I_{e} = I_{e} + F_{size}$}
      \STATE{$Frame = Frame + 1$}
            
      \ENDWHILE
      \RETURN $Segmented\_sliding\_windows$
      \end{algorithmic}
 \end{algorithm}
\vspace{-.9em}

%
%
\subsection{Data Preparation for Convolutional Neural Network}
    \label{section:dataPrepCNN}
    After pre-processing of signals, we prepared $205$ sessions (each $40$ seconds long) of data at different walking or running speeds. Both tri-axial accelerometer and gyroscope sensors were sampled at $51$ Hz which makes total dataset dimension to $[418200, 6]$. Next, we segmented the data into $3$ seconds moving windows with $50\%$ overlapping as described in Algorithm \ref{label:slidingWindows}. It converts the dataset to $[5465,153,6]$ dimension where total input samples are $5465$ and each input dimension is $[153,6]$. We decided on the window size and the overlapping percentage based on our prior experience from the studies in \cite{Bharti2017, bharti2018human}.

%

\vspace{-.5em}

\subsection{Neural Network Architecture}
\label{section:ArchitectureCNN}

    Convolution neural networks (CNN) are competent in automatically extracting the relevant features from the input signals. It has achieved promising results in solving image classification, speech recognition, and text analysis problems \cite{lecun2015deep}. In this work, we wanted our CNN model to learn from the accelerometer and gyroscope data independently at early stages and later merge these information to predict the final speed. To do so, we designed a CNN architecture (as shown in Figure~\ref{fig:speedArchitecture}) that has $2$ symmetrical branches. First branch takes its input from the accelerometer data and second from the gyroscope in the input layer. Input layer is followed by the convolutional layers in each branch. To optimize the hyper-parameters, we performed a randomized search on the the range of following values - number of convolution layers ($2-10$), number of filters in the convolution layers ($10-100$), number of dense layers ($2-5$) and number of neurons in each dense layer ($15-500$). We randomly sampled $50$ settings of hyper-parameters and kept the one which performed best. Our hyper-parameter search led us to add two convolutional layers - the first layer consists of $27$ filters and second $45$ filters, where each filter size is $3\times 3$. These convolution layers extract both high-level and low-level representations of the data. The output of these convolutional layers is passed to a global max-pooling layer which distills the output of the convolution filters into a salient vector of size $45$ by extracting the maximum value from each filter. After pooling, the outputs of accelerometer and gyroscope convolutional layers are concatenated to fuse the information from both layers which produced a vector of size $90$. Finally, this vector is fed to the two sequential dense layers. The two dense layers have $180$, and $30$ neurons respectively. Output of second dense layer is forwarded as an input to the single neuron output layer that estimates the final speed. 

\subsection{Training procedure of CNN model}
    \label{section:trainCNN}
We employed two strategies to train our CNN model. In the first strategy, we divided the dataset in three subsets - train, test and evaluation, which were used for training, fine-tuning and final evaluation for the model respectively. In the second strategy, we employed leave-one-out cross-validation method where model was trained on the data from $14$ participants and evaluated against the data from remaining $1$ participant. We cycled this process $15$ times so that once each participant data was included in the evaluation. Finally, we computed the mean to achieve the final results. 

For both of these strategies, the training of CNN model consists of $3$ main processes. In the first process called feedforward propagation, data propagates from the input layer to the output layer. In this process, model makes a prediction of speed based on the input sensory data. In the second process called loss function, model computes the prediction loss which represents how far is the predicted speed off from the real speed. In the third process called backpropagation, model adjust its trainable parameters such that it predicts the speed closer to the real one. While training, our CNN model went through all the $3$ processes in each epoch to optimize the parameters. We used mean absolute error (MAE) as a loss function to compute the difference between predicted and real speed. Further, we employed RMSprop optimizer algorithm which is similar to the gradient descent with momentum. It restricts the oscillations of loss function in the vertical direction, which helps to increase the learning rate and as a result enables the model take larger steps in the horizontal direction to converge faster. Additionally, to make sure our model is not suffering from the overfitting problem, we leveraged Dropout regularization technique that makes the training process noisy by randomly dropping the neurons within a layer such that one neuron doesn't take too much responsibility to make the final prediction. We trained our model to $1000$ epochs and used early stopping criteria that stops the training if the MAE is not decreasing for $10$ consecutive epochs.

\vspace{-1em}
\begin{figure}[!ht]
    \centering
  {\includegraphics[trim=5 5 5 5,clip,width = \linewidth, keepaspectratio]{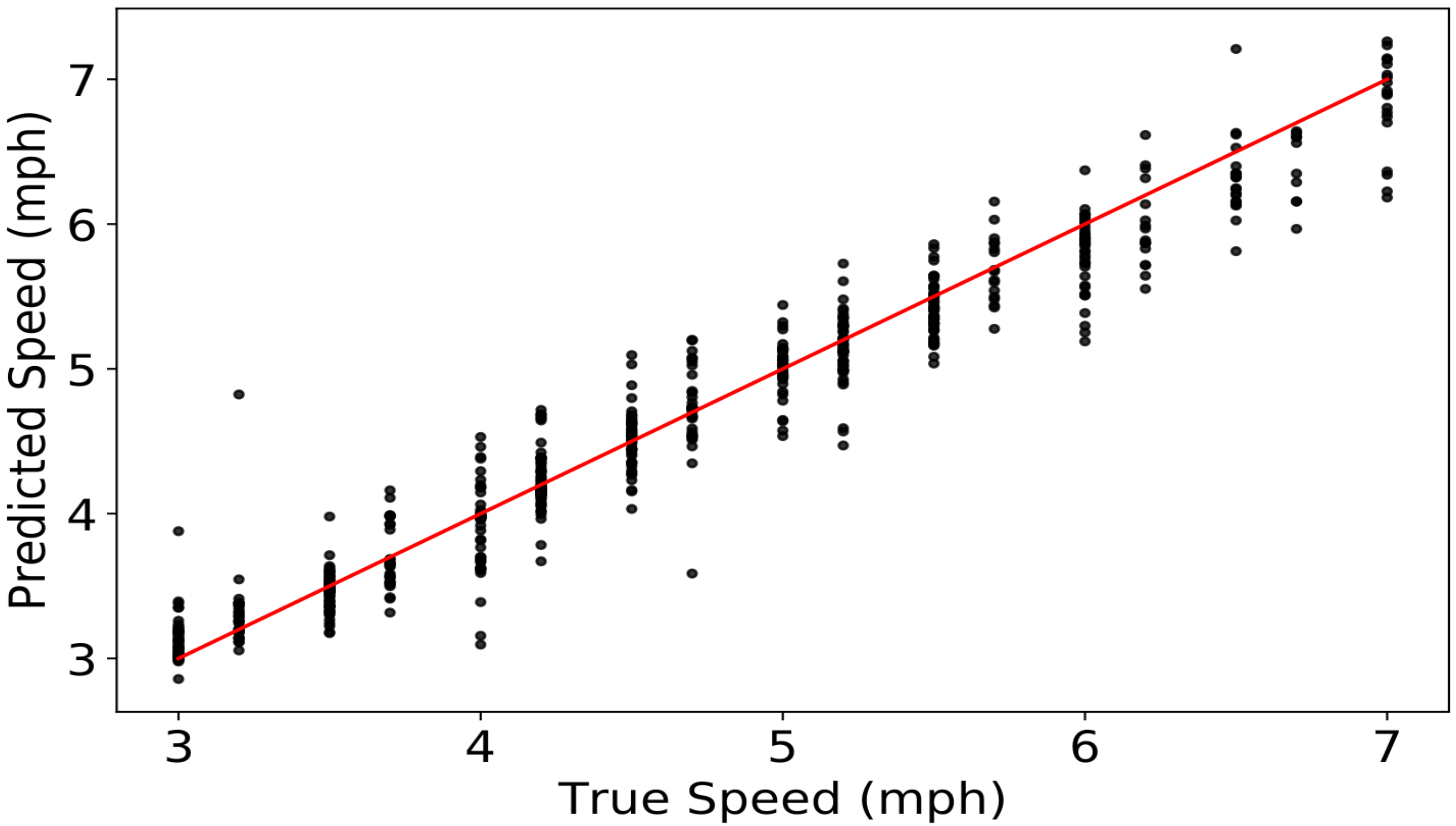}}
 {\includegraphics[trim=5 5 5 5,clip,width = \linewidth, keepaspectratio]{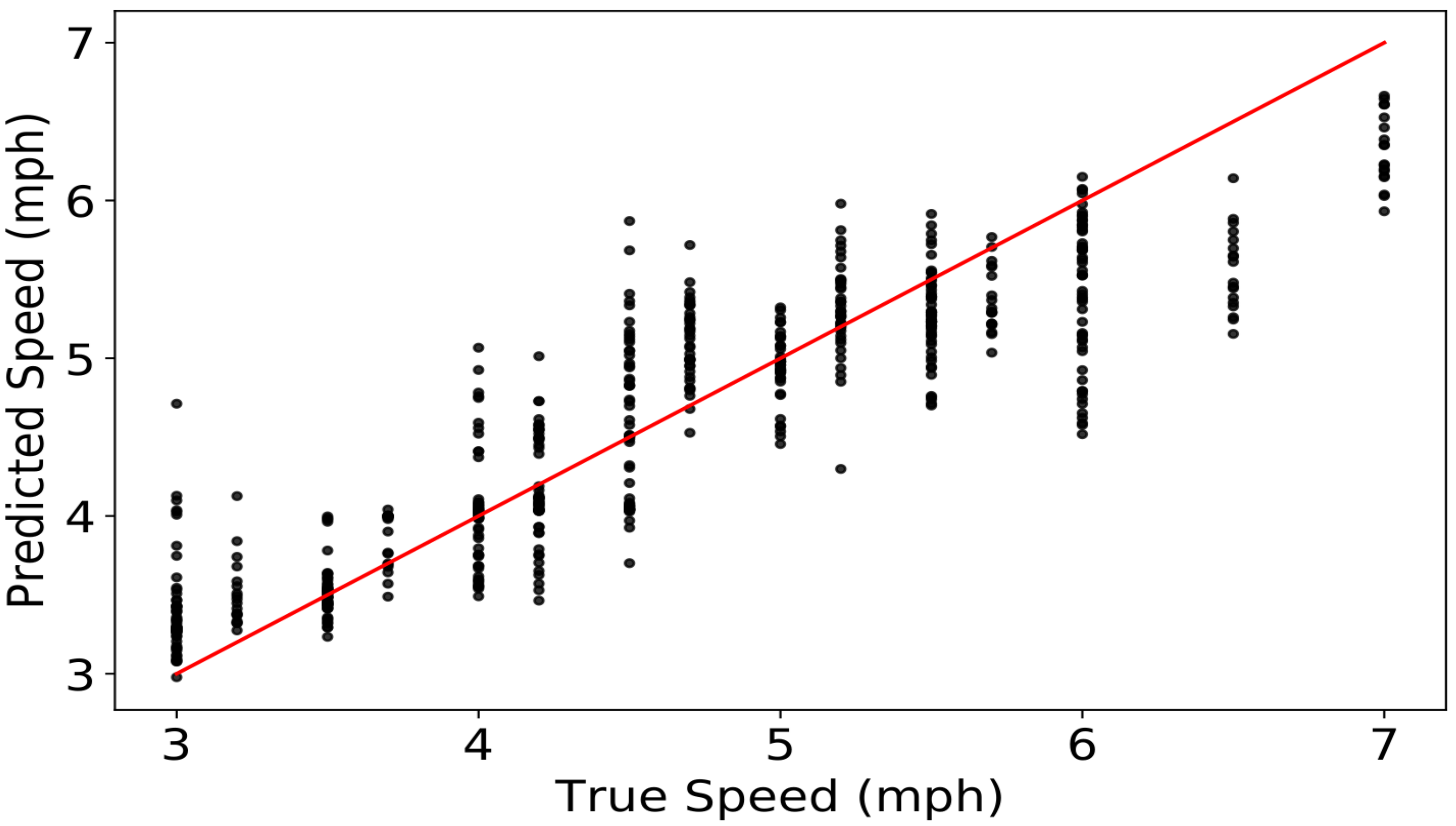}}
  \caption{True vs predicted speeds for train-test-evaluation split (top) and leave one participant out cross validation (bottom)}
    \label{fig:predictionsScatter}
\end{figure}
\vspace{-1.5em}

\section{Results}
\label{section:Results}

    We evaluated our speed detection model against $3$ evaluation metrics - Mean Absolute Error (MAE), Mean Absolute Percentage Error (MAPE), and R-Squared ($R^2$) as described in equation \eqref{eq:MAE}, \eqref{eq:MAPE} and \eqref{eq:RSquare} respectively. MAE measures the average magnitude of the error in the prediction, without considering its direction. It uses the same scale as the data and has no bounded range, hence can not be interpreted without knowing the scale of real values. In contrast, MAPE measures the percentage error in the prediction based on the predicted and real value. In comparison to MAE, MAPE can be interpreted well because its range is bounded between $0$ and $100$. $R^2$ is based on the ratio of the prediction error and the variance in the predicted values. It ranges between $0$ and $1$, where $0$ indicates that the model explains none of the variability of the data around its mean and $1$ indicates it explains all the variability.  If $S_{i}$, $\hat{S_{i}}$ and $\bar{S_{i}}$ are predicted, real and mean predicted speeds respectively, then


\vspace{-0.5em}
\begin{equation}\label{eq:MAE}
    MAE=\dfrac{1}{n}\sum_{i=1}^{N}|S_{i}-\hat{S_{i}}|
\end{equation}
\begin{equation}\label{eq:MAPE}
    MAPE=\dfrac{1}{n}\dfrac{\sum_{i=1}^{N}|S_{i}-\hat{S_{i}}|}{S_{i}}
\end{equation}
\begin{equation}\label{eq:RSquare}
    R^2=1-\dfrac{\sum_{i=1}^{N}(S_{i} - \hat{S_{i}})^2}{\sum_{i=1}^{N}(S_{i} - \bar{S_{i}})^2}
\end{equation}
\vspace{-1.75em}

\begin{figure}[!ht]
    \centering
    \includegraphics[scale = 0.5, keepaspectratio]{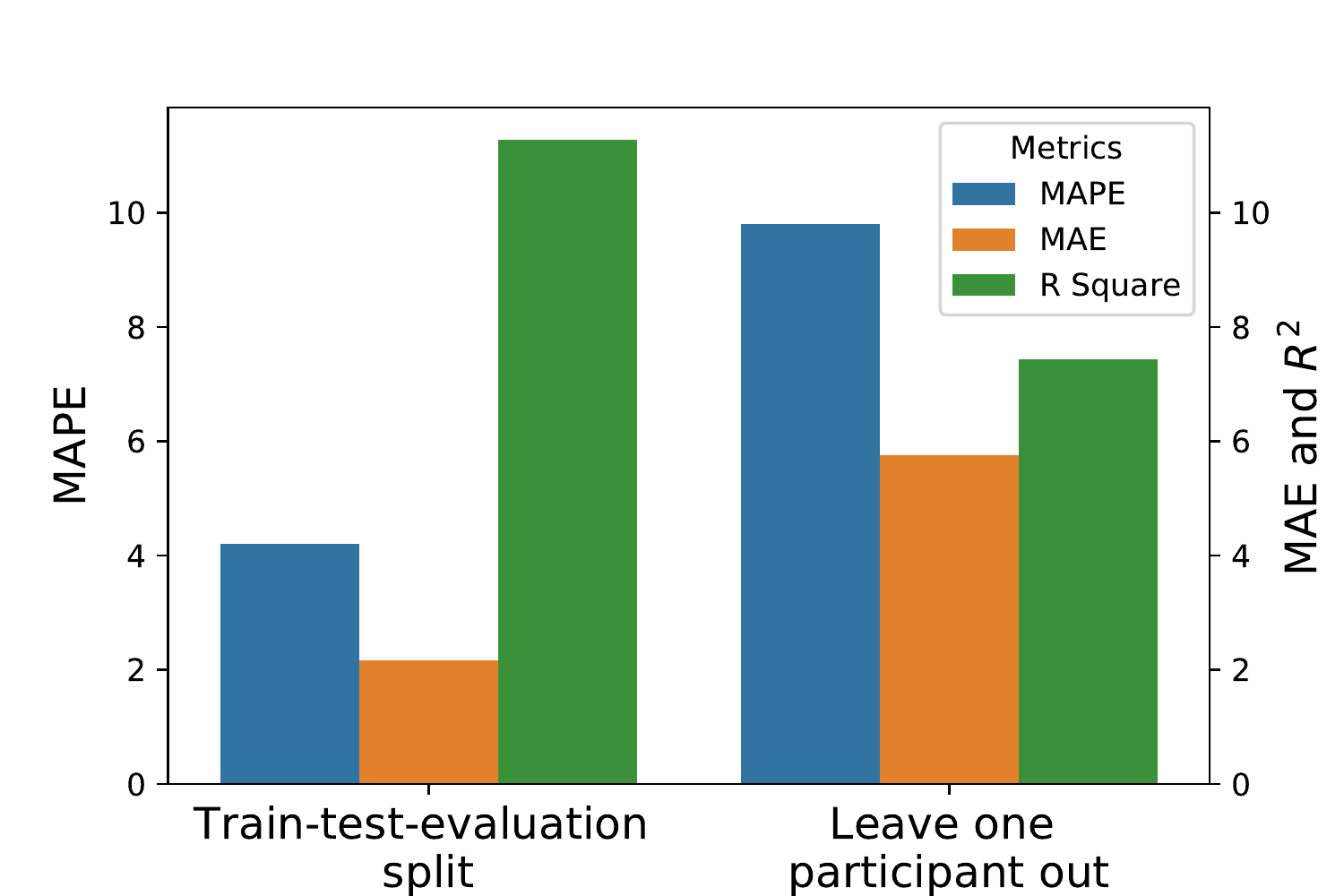}
    \caption{Results for speed estimation algorithm against train-test-evaluation split and leave-one-out evaluation strategies}
    \label{fig:speedEstimationResults}
\end{figure}

    For train-test-evaluation split evaluation strategy, our model achieved $0.18$, $4.2\%$, $0.94$ value of MAP, MAPE and $R^2$ respectively as shown in Figure~\ref{fig:speedEstimationResults}. In this evaluation strategy, same participant data can be present in train, test and evaluation subset which can induce bias in the results. To avoid this, we evaluated our model on more stringent strategy, leave-one-out cross-validation. For this strategy, model achieved $0.48$, $9.8\%$, and $0.62$ value of MAP, MAPE and $R^2$ respectively. As shown in Figure \ref{fig:predictionsScatter}, for both strategies predictions are  fitted close to the regression line at different speeds. At higher speed, model has high error rate for leave-one-out strategy. It is likely because each participant exhibits different running posture at the higher speed. Since the participant data which is used for evaluation is not included in the training, it is difficult to capture those patterns. We applied model quantization techniques to optimize the model and make it compatible for low end devices. 
\vspace{-.5em}
\section{Conclusion and Discussion}
\label{section:Discussion}
Our speed detection model based on CNN regressor architecture has achieved favorable results against train-test-evaluation split and leave-one-out cross-validation strategies. Higher accuracy of our solution makes it particularly useful to the elder citizens who are advised to be as active as  possible and an error-prone device can discourage them from exercising. Unlike GPS, our IMU sensor based wearable solution can work well at indoors as well as outdoors. Since, step-frequency doesn't change  significantly when speed increases within the walking or running,  it becomes difficult to improve the accuracy. To overcome this challenge, in future study, we will design a $2$-phase speed detector model, where in first phase, we will classify running vs walking and then, in second phase will estimate the speed withing that activity.

\bibliographystyle{./bibliography/IEEEtran}
\bibliography{./bibliography/IEEEabrv,./bibliography/IEEEexample}
\end{document}